\documentclass[12pt]{article}

\usepackage{epsf}
\usepackage[dvips]{graphicx}
\begin{document}
\begin{center}
{\bf\large{Exotic narrow resonance searches in the system $\Lambda
K^0_s$ in p+propane collisions at 10 GeV/c}}

\vskip 2mm P.Zh.Aslanyan$^{1,2 \dag}$ , V.N.Emelyanenko$^1$, G.G.
Rikhkvitzkaya$^1$

\vskip 5mm {\small (1) {\it Joint Institute for Nuclear Research }
\\
(2) {\it Yerevan State University }
\\
$\dag$ {\it E-mail: paslanian@jinr.ru }}

\end{center}

\begin{abstract}
Experimental data from the 2m propane bubble chamber have been
analyzed to search for an exotic baryon states, in the $\Lambda
K^0_s$ decay mode for the reaction p+$C_3H_8$ at 10 GeV/c. The
invariant mass spectrum $\Lambda K^0_s$ observe a narrow peaks at
1750$\pm$18, 1795$\pm$18,1850$\pm19$ MeV/$c^2$ and full widths of
 $\Gamma_{exp.}$= 32$\pm$6, 44$\pm$15,
29.0$\pm$8 MeV/$c^2$. The statistical significance of these peaks
has been estimated as 5.6, 3.3 and 3.0 S.D., respectively. There are
the small enhancements in  mass regions of (1650-1675) and
(1925-1950) ÌýÂ/ñ$^2$. These would be  candidates for the $N^0$ or
the $\Xi^0$ pentaquark states.

The investigation has been performed at the Veksler and Baldin
Laboratory of High Energies, JINR.

\end{abstract}

\section{introduction}
Several models predict the multiplet structure and characteristics
of pentaquarks for example the chiral soliton model, the
uncorrelated quark model, correlated quark models, QCD sum rules,
thermal models, lattice QCD etc\cite{1}-\cite{18}. Multi-quark
states, glueballs and hybrids have been searched for experimentally
for a very long time, but none is established.

 Results from a wide range of recent experiments\cite{19}
are consistent with the  existence of an exotic S=+1 resonance, the
$\Theta+(1540)$ with a narrow width and a mass near 1540 MeV [1].
Results from this experiment: $M_{\Theta^+}$ = (1540$\pm$8)
MeV/$c^2$, $\Gamma_{\Theta^+}$ =(9.2$\pm$1.8) MeV/$c^2$(PDG-04:
$\Gamma_{\Theta^+}$ =(9.2 $\pm$0.3) MeV/$c^2$).

However, recent significant advances in theoretical  and
experimental work led to a number of new candidates in the last 2
years of searches. Candidates for other pentaquarks have been
presented recently, in particular for the
$\Xi^{--}(1862)$,$\Xi^-(1850)$, $\Xi^0(1864)$\cite{20} and the
$\Theta^0_c(3099)$ \cite{21}. Preliminary results of the STAR
(Solenoidal Tracker At RHIC) on a search for the $\Xi^0$ I=1/2 as
  well as for the $N^0$ or the $\Xi^0$ pentaquark states in the decay mode $\Lambda K^0_s$ with the mass
  $1734\pm0.5\pm5$ MeV/$c^2$  is presented in the article\cite{22}.  A significant
signal for $\approx \Xi^0(1750)\to \Xi^- \pi^+$ was
observed\cite{23}.

\section{Experiment}

The JINR 2m bubble chamber is the most suitable instrument for this
purpose \cite{24}. The experimental information of more than 700000
stereo photographs are used to select the events with $V^0$ strange
particles.

 The events with $V^0$ ($\Lambda$ and $K^0_s$)  were identified
by using the following criteria \cite{24}:\\
 1) $V^0$ stars from the photographs were selected
according to $\Lambda\to\pi^-+p$, neutral $K_s\to\pi^-+\pi^+$ or
$\gamma \to e^++e^-$ hypothesis. A momentum limit of $K^0_s$ and
$\Lambda$ is  greater than 0.1 and 0.2 GeV/c, respectively ; 2)
$V^0$ stars should have the effective mass of $K^0_s$ and of
$\Lambda$; 3) these $V^0$ stars are directed to  some
vertices(complanarity); 4) they should have one vertex, a three
constraint fit for the $M_K$ or $M_{\Lambda}$ hypothesis  and
after the fit, $\chi^2_{V^0}$ should be selected  over range less
than 12; 5) The analysis has shown\cite{24} that the events with
undivided $\Lambda, K^0_s$ were appropriated events as
$\Lambda$

The effective mass distribution of 8657-events with $\Lambda$,
4122-events with $K^0_s$ particles are consistent with their PDG
values(Fig.1). The effective mass resolution of $\Lambda K^0_sp$
system  was estimated to be on the average 1\%.

 Each $V^0$ event weighted by a factor $w_{geom}$ (=1/$e_{\tau}$), where $e_{\tau}$
is the probability for potentially observing the $V^0$, it can be
expressed as
$$
e_{\tau}= exp(-L_{min}/L)- exp(-L_{max}/L),
$$
where L(=cp$\tau$/M) is the flight length of the $V^0$,$L_{max}$
the path length from the reaction point to the boundary of
fiducial volume, and $L_{min}$(0.5 cm) an observable  minimum
distance between the reaction point and the $V^0$ vertex.
M,$\tau$, and p are the mass, lifetime, and momentum of the $V^0$.
The average geometrical weights were 1.34$\pm$0.02 for $\Lambda$
and 1.22$\pm$0.04 for $K^0$.  Figure 2 compares the momentum,
 cos$\theta$ in the c.m. nucleon-nucleon system, transverse momentum($p_t$) and
 longitudinal rapidity distributions  of $\Lambda$ and $K^0_s$
 for experimental events (solid line) and those simulated by the FRITIOF
model\cite{26,27} (broken line)in p+C interactions. From Fig.2 one can see that
the experiment is satisfactorily described by the FRITIOF model.

 The estimation of experimental inclusive cross
sections for $\Lambda$ and $K^0_s$ production in the p$^{12}C$
collision is equal to $\sigma_{\Lambda}$= 13.3$\pm$1.7 mb and
$\sigma_{K^0_s}$= 3.8$\pm$0.6 mb, respectively \cite{24}.

\section{$\Lambda K^0_s$ - spectrum  analysis}

The total experimental background has been  obtained by three
methods(Fig.3). In the first method, the experimental effective mass
distribution was approximated by the polynomial function  after
cutting out the resonance ranges because this procedure has to
provide the fit with $\chi^2$=1 and polynomial coefficient with
errors less than 30\%. This distribution was fitted by the six-order
polynomial. The second of the randomly mixing method of the angle
between  $K^0_s$  and $\Lambda$  for experimental events is
described in \cite{25}. Then, these background events were analyzed
by using the same experimental condition and the effective mass
distribution ( $\Lambda K^0_s$ )  was fitted by the six-order
polynomial. The third type of background for ($\Lambda K^0_s$ )
combinations has been obtained by FRITIOF model\cite{26,27}(Fig.3b).
In all figures the background distribution  has been normalized to
the experimental distribution . The analysis of background done by
three methods has shown that there are not observable structure  in
range of peaks. The analysis of the experimental data are based on
the polynomial method.

 Figure  3  shows  the invariant mass of
1012 ($\Lambda K^0_s$ )combinations with bin sizes 10 MeV/$c^2 $.
The values for the mean position of the peak and the width obtained
by using Breit Wigner fits. There are significant enhancements in
mass regions of 1750,1795 and 1835 MeV/$c^2$(Fig.3). Their excess
above background by the first method is 4.0, 2.7 , 3.0 S.D.. There
is small enhancement in mass region of 1935 MeV/$c^2$. The
simulation with FRITIOF model for ($\Lambda K^0_s$ )combinations has
shown in Fig.4 that there are not significant reflection from well
known resonances in this distribution. Similar results have obtained
when using a Breit Wigner distribution and different bin sizes.

Figure 5  shows  the invariant mass of ($\Lambda K^0_s$ ) with bin
sizes 11 MeV/$c^2 $.There are significant enhancements in mass
regions of 1670,1750,1795 and 1850 MeV/$c^2$(Fig.5). Their excess
above background by the first method is 2.9, 4.7 , 2.3 and 2.4 S.D..

 The effective mass distribution of ($\Lambda K^0_s$ ) with bin size 18
MeV/$c^2$ is shown in Fig.6. This bin size is consistent with the
experimental resolution within the errors.The solid curve is the sum
of the background by the first method and  2 Breit-Wigner resonance
curves(Fig 6). There are significant enhancements in mass regions of
1750 and 1795 MeV/$c^2$. Their excess above background by the first
method is 5.6  and  3.3 S.D..respectively. There are negligible
enhancements in the mass regions of 1680, 1860 and 1950 MeV/$c^2$.

 \section{Conclusion}

 A number of peculiarities were found in the effective mass spectrum
of system $\Lambda K^0_s$ in ranges of:(1740-1750), (1785-1800) and
(1835-1860) MeV/$c^2$ in collisions of protons of a 10 GeV/c
momentum with propane nuclei. The detailed research of structure of
mass spectrum has shown, that the maximum statistical significance
has been obtained in effective mass ranges submitted in table 1.
There are small enhancements in  the mass spectrum regions of
(1650-1675) and (1925-1950) MeV/$^2$.

The preliminary total cross section for N0(1750) production in
p+$C_3H_8$ interactions is estimated  to be $\approx 30 \mu$b.

 The $N^0$ can be  from  the antidecuplet, from an octet (D. Diakonov, V. Petrov,
 \cite{1}, V.Guzey and M.Polyakov,\cite{2}) or an  27-plet(J. Ellis et al, \cite{8}). On the other hand,
 Jafe and Wilczek  predicted a mass around 1750 MeV and a width 50 \% larger for
 these states than that of the $\Theta^+$(\cite{5}).

 These peaks are  possible candidates for two pentaquark states:
  the $N^0$ with quark content udsds decaying into $\Lambda K^0_s$ and the $\Xi^0$  quark content
  udssd decaying into $\Lambda \overline{K^0_s}$, which  are agreed:
  with the calculated rotational spectra $N^0$  and $\Xi^0$ spectra from  the theoretical report  of
   D. Akers,\cite{28}(Fig.7) and with $\Theta^+$
   spectra from  the experimental  reports  of  Yu.A.Troyan et.al.,JINR,
    D1-2004-39, Dubna,2004  and  P. Aslanyan JINR, E1-2004-137,2004.

\begin{table}
\caption{The statistical significance, the width($\Gamma$) and the effective mass resonances
in collisions of protons with propane at 10 GeV/c}
 \label{tab:res}
\begin{tabular}{|c|c|c|c|c|c|c|c|}  \hline
Resonance & $M_{\Lambda K^0_s}$&Experimental&&The maximal  \\
Decay & MeV/$c^2$&width $\Gamma_e$&$\Gamma$&statistical\\
Mode & &MeV/$c^2$&&significance $N_{sd}$\\ \hline
  $K^0_s\Lambda$&1750$\pm$18&32$\pm$6&14$\pm$6&5.6\\
 $K^0_s\Lambda$&1795$\pm$18&44$\pm$15&26$\pm$15&3.3\\
$K^0_s\Lambda$&1850$\pm$19&28$\pm$7&10$\pm$7&3.0\\
\hline
\end{tabular}
\end{table}


\begin{figure}[ht]
 \epsfysize=150mm
 \centerline{
 \epsfbox{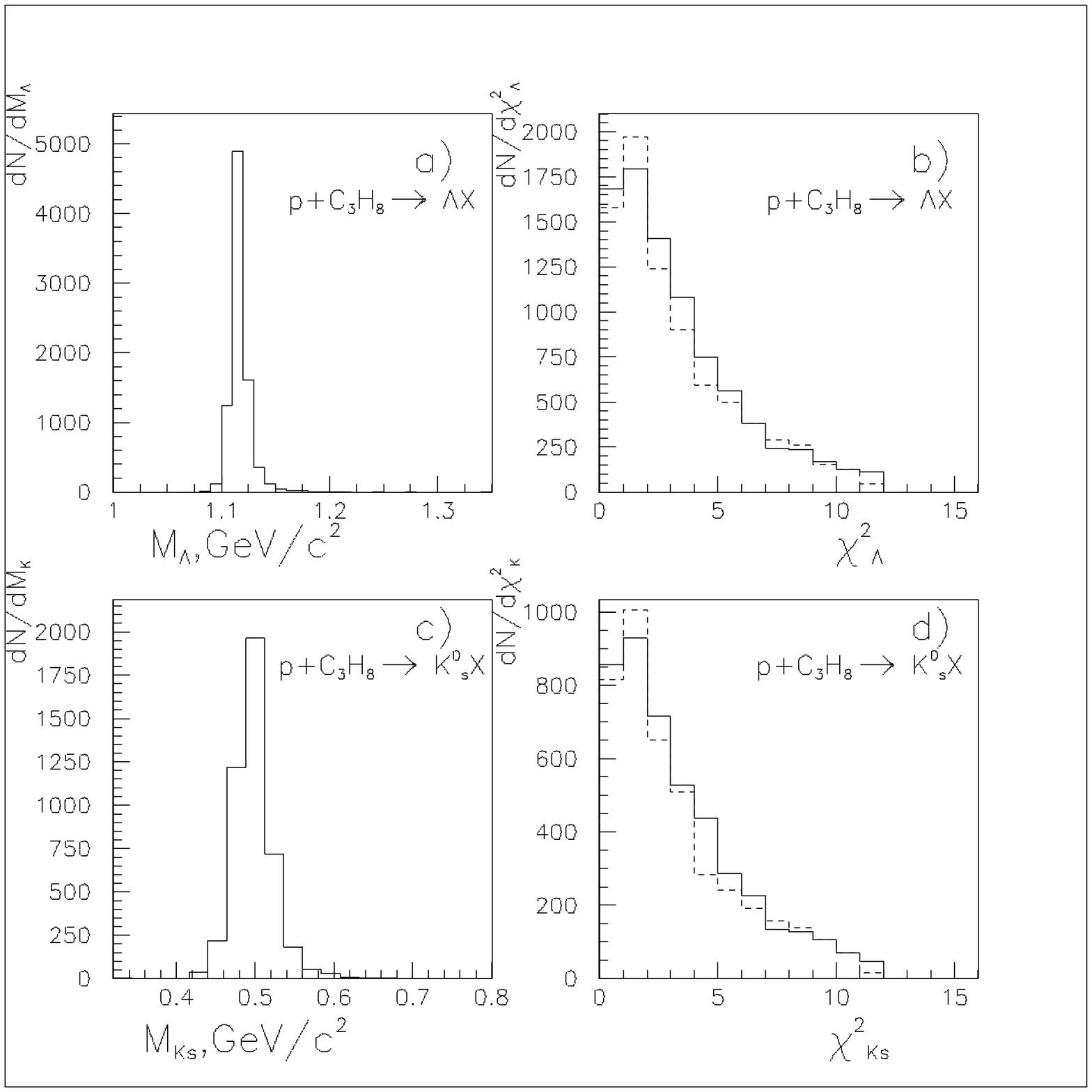}}
 \caption{ The distribution of experimental $V^0$ events produced
 from interactions of beam protons with propane: a) for the effective mass of
  $M_{\Lambda}$; b)for $\chi^2_{\Lambda}(1V-3C)$ of the fits via the decay mode
$\Lambda\to \pi^-+p$; c) for the effective mass of
$M_{K^0_s}$;d)for $\chi^2_{K^0_s}(1V-3C)$ of the fits via decay
mode $K^0_s\to\pi^-+\pi^+$. The expected functional form for
$\chi^2$ is depicted with the dotted histogram.}
\end{figure}

\begin{figure}[ht]
 \epsfysize=180mm
 \centerline{
 \epsfbox{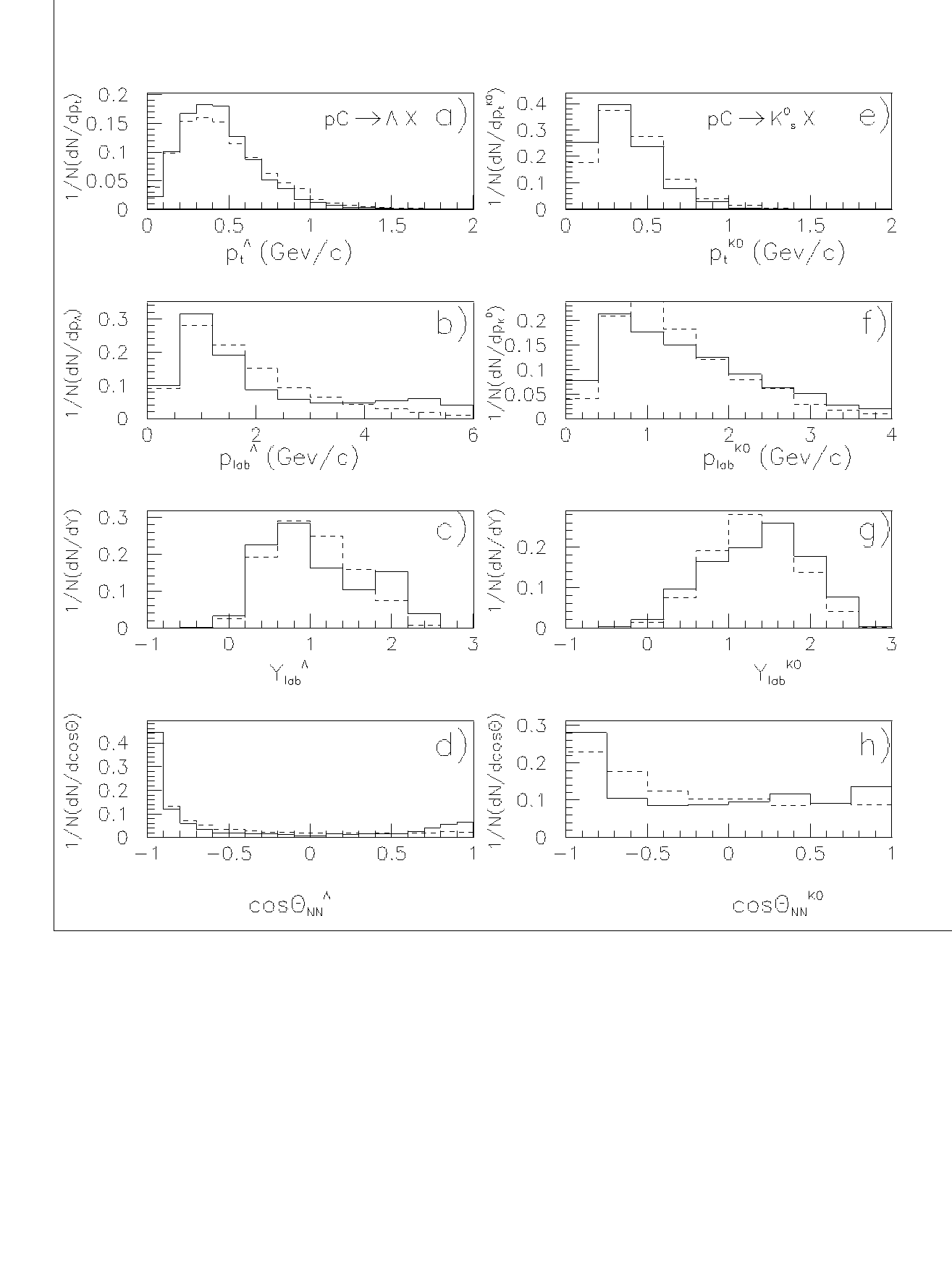}}
 \caption{Experimental(solid) and simulation by FRITIOF
model(dashed)  distributions of $\Lambda$ -  hyperons and $K^0_s$-
mesons in p+C interaction at 10 GeV/c: a)and e) by the transverse
momentum ($p_t$); b) and f)by the momentum ($p_{lab}$); c) and g) by
the longitudinal rapidity ($Y_{lab}$); d)and h) by the azimuthal
angle cos$\Theta$(in the SM of p+p collisions.  }
\end{figure}

\begin{figure}[ht]

 \centerline{\epsfysize=180mm
 \epsfbox{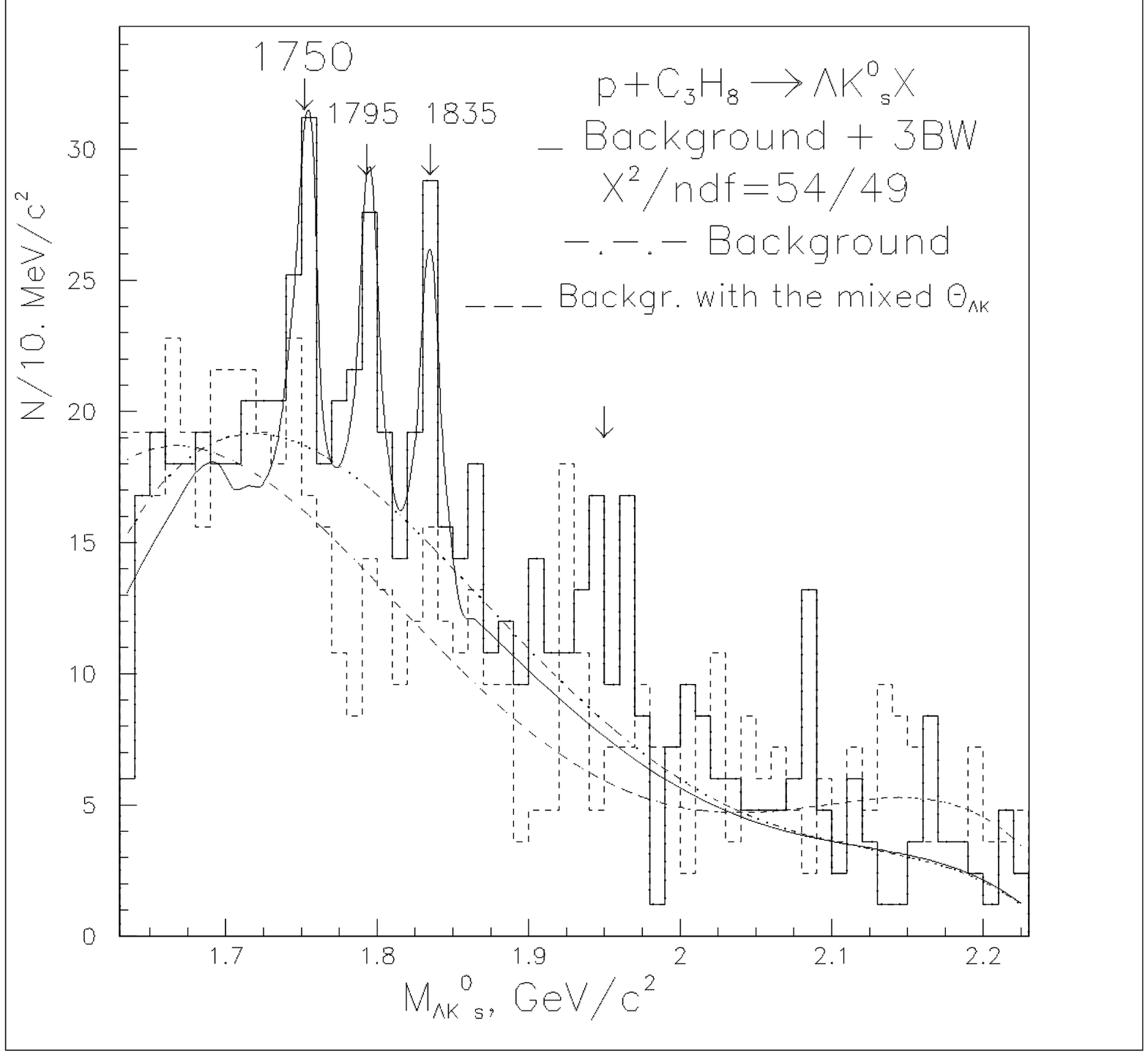}}
 \caption{Invariant mass distribution ( $\Lambda K^0_s$) with the bin size 10
 MeV/$c^2$ in the inclusive reaction p+$C_3H_8$ . The solid curve is the sum of the
 experimental background by the first method (the dot-dashed curve)  and
  3 Breit-Wigner  resonance curves. The dashed histogram is the
  experimental background[25].  }
  \end{figure}

  \begin{figure}[ht]
 \epsfysize=180mm
 \centerline{
 \epsfbox{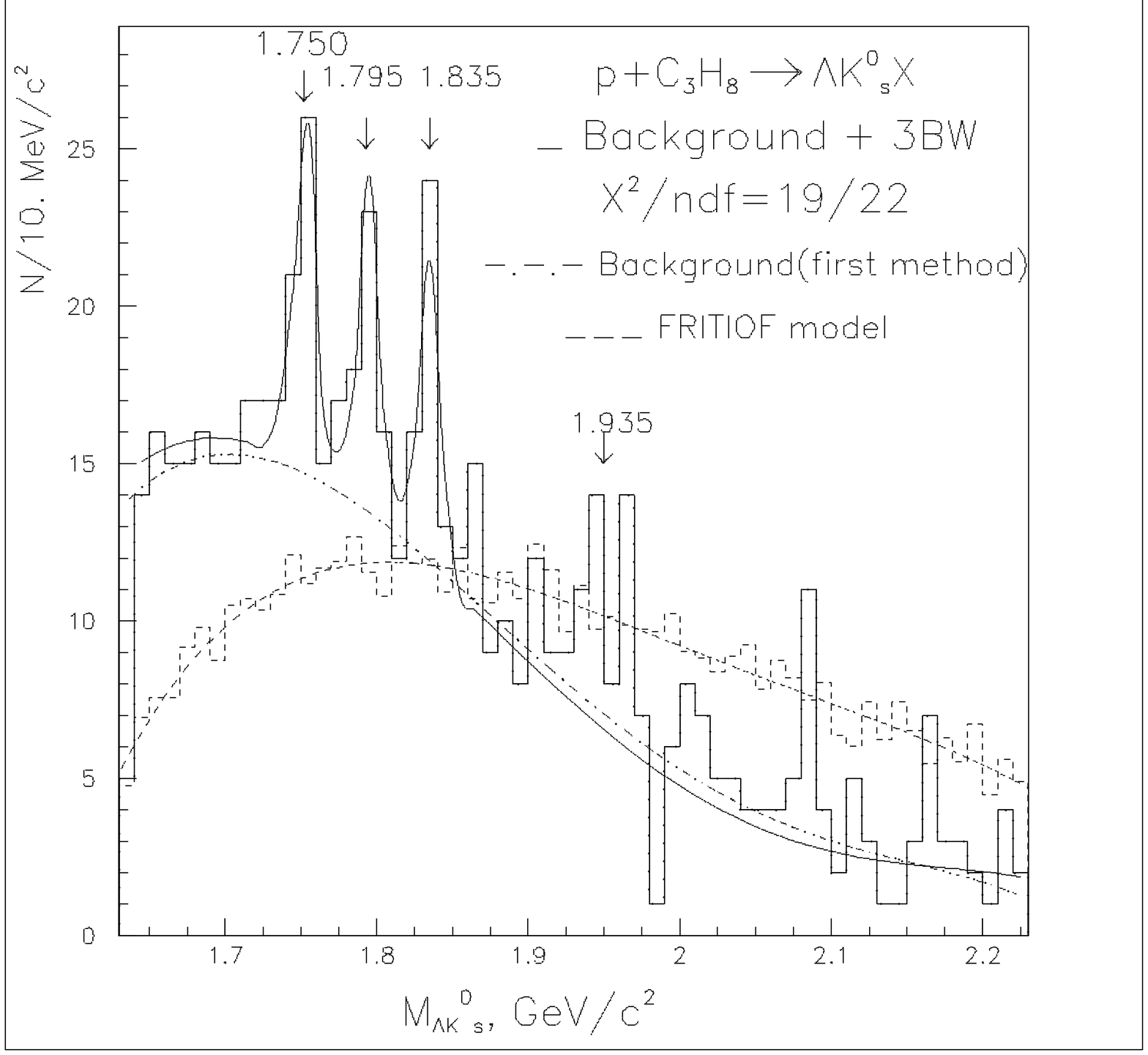}}
 \caption{Invariant mass distribution ( $\Lambda K^0_s$)with the bin size 10
 MeV/$c^2$ in the inclusive reaction p+$C_3H_8$ . The solid curve is the sum of the
 experimental background by the first method (the dot-dashed curve)  and
  3 Breit-Wigner  resonance curves. The dashed histogram is the
  simulation by FRITIOF model[26,27] .
  }
\end{figure}

   \begin{figure}[ht]
 \epsfysize=180mm
 \centerline{
 \epsfbox{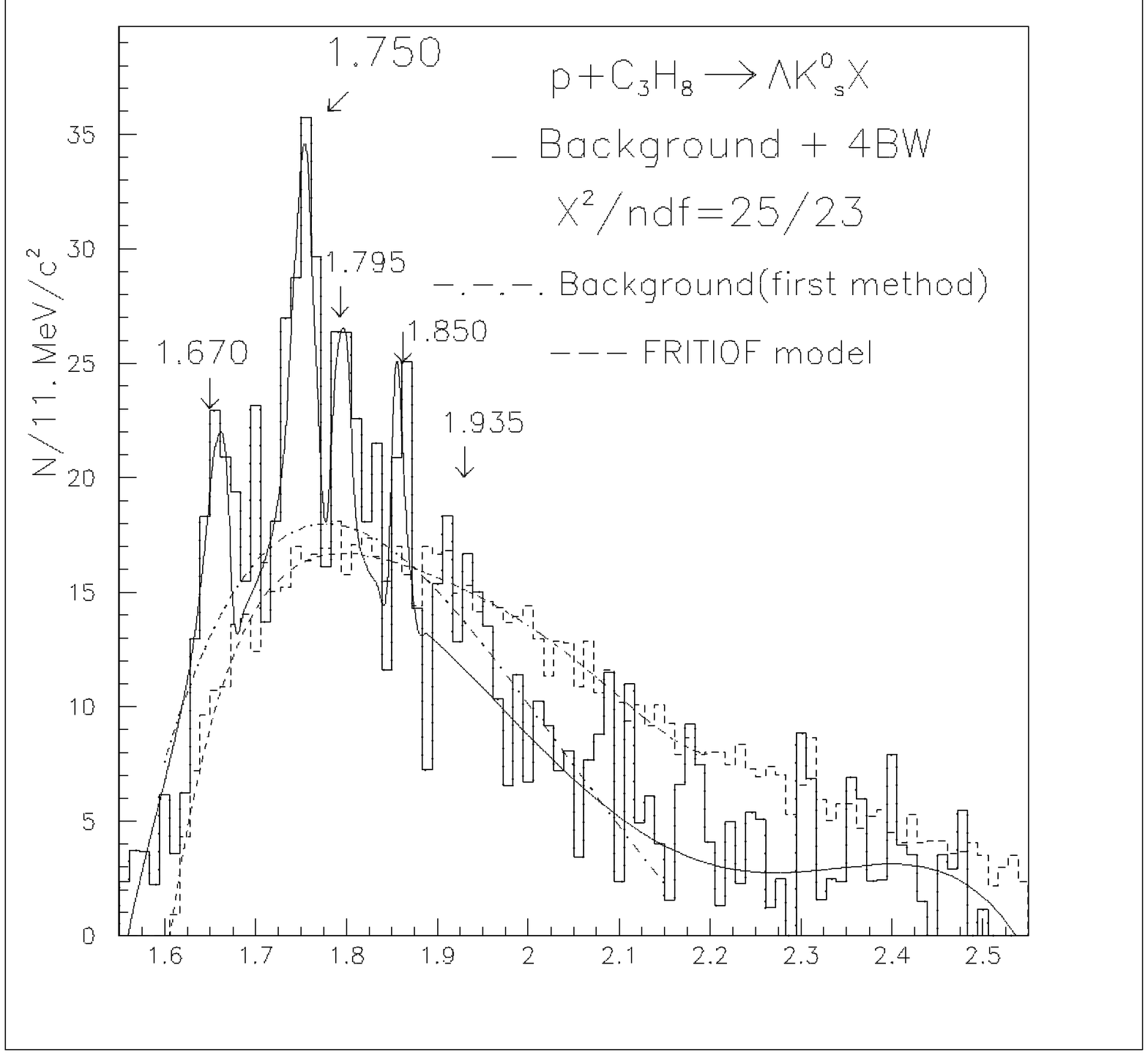}}
 \caption{Invariant mass distribution ( $\Lambda K^0_s$) with the bin size 11
 MeV/$c^2$ in the inclusive reaction p+$C_3H_8$ . The solid curve is the sum of the
 experimental background by the first method (the dot-dashed curve)  and
  4 Breit-Wigner  resonance curves. The dashed histogram is the
 simulation by FRITIOF model[26,27].
  }
\end{figure}

\begin{figure}[ht]
 \epsfysize=180mm
 \centerline{
 \epsfbox{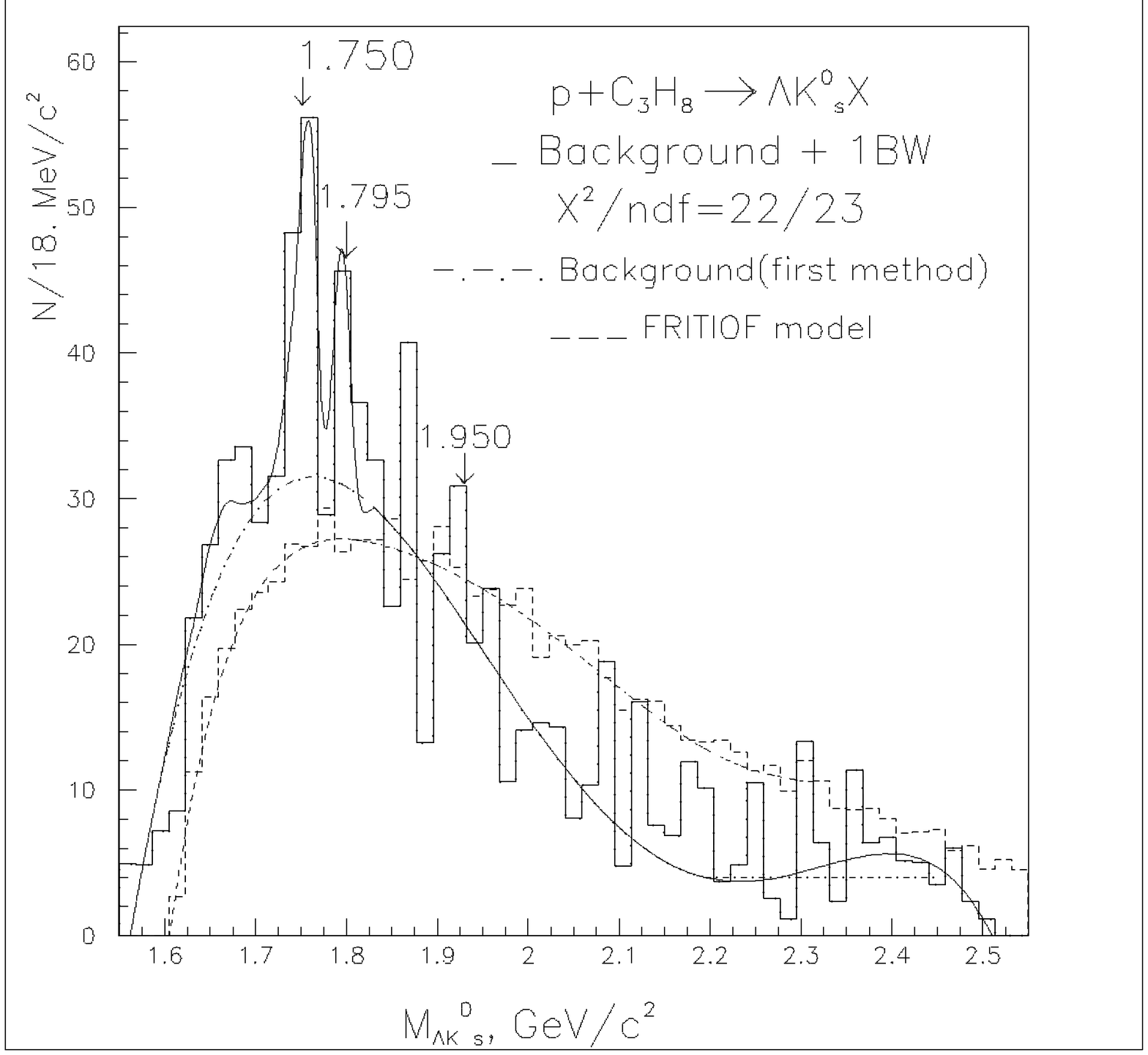}}
 \caption{Invariant mass distribution ( $\Lambda K^0_s$)with the bin size 18
 MeV/$c^2$ with the bin size 10in the MeV/$c^2$ in the inclusive reaction p+$C_3H_8$ . The solid curve is the sum of the
 experimental background by the first method (the dot-dashed curve)  and
  2 Breit-Wigner  resonance curves. The dashed histogram is the
 simulation by FRITIOF model[26,27].
  }
\end{figure}

\end{document}